\documentclass[preprint,showpacs,showkeys,12pt]{revtex4-1}
\usepackage{ifpdf}
\ifpdf
\usepackage{hyperref}
\else
\usepackage[hypertex]{hyperref}
\fi
\usepackage{graphicx,xcolor,subfigure}
\usepackage{amsmath,amsfonts,amssymb,pb-diagram}
\usepackage[nodayofweek]{datetime}

\newcommand{\f}{\frac}

\begin{document}

\title{Physical interpretation of point-like interactions of one-dimensional Schr\"{o}dinger operator}

\author{V.~L. Kulinskii, D.~Yu. Panchenko}
\email[V.~L. Kulinskii: ]{kulinskij@onu.edu.ua}
\email[D.~Yu. Panchenko: ]{dpanchenko@onu.edu.ua}

\affiliation{Department of Theoretical Physics, Odessa National University, Dvoryanskaya 2, 65082 Odessa, Ukraine}

\begin{abstract}
We consider physical interpretations of non-trivial boundary conditions of self-adjoint extensions for one-dimensional Schr\"{o}dinger operator of free spinless particle.
Despite its model and rather abstract character this question is worth of investigation due to application for one-dimensional nanostructures. The main result is the physical interpretation of peculiar self-adjoint extension with discontinuity of both the probability density and the derivative of the wave function. We show that this case differs very much from other three which were considered before and corresponds to the presence of mass-jump in a sense of works of Ganella et. al. \cite{qm_deltamassjump_physlett2007,qm_deltamassjump_jphysmath2009} along with the quantized magnetic flux. Real  physical system which can be modeled by such boundary conditions is  the localized quantazied flux  in the Josephson junction of two superconductors with different effective masses of the elementary excitations.
\end{abstract}


\maketitle


\section*{Introduction}
Current development of nanoengineering allows to construct quantum system with controllable properties. Such systems  may be constructed so that its design closely correspond to exactly solvable models. The existence of localized defects in nanostructures can be modeled by the point-like interactions. The method of zero-range potential is widely used in situations where characteristic scale of the interaction much smaller that the wave length of the particles \cite{qm_delta0fermi_1936} (see also \cite{book_bazeldovichperelomov_en,book_demkovostrvsk_en}). Mathematical framework which gives unified way to obtain possible boundary conditions for such singular interactions is the theory of self-adjoint extensions of hermitian operators \cite{qm_deltaberezinfaddeev_dan1961en,math_book_selfadjext}.
But such an approach does not provide the understanding of regularization one needs to perform in order to get the limiting boundary conditions representing singular interaction. Surprisingly, this happens even in the simplest case of free particle hamiltonian:
\begin{equation}\label{freeham}
\hat{H}_{0} = -\f{d^2}{d\,x^2}\,.
\end{equation}
on the axis $x\in (-\infty; +\infty)$ which has deficiency indices $(2,2)$ and consequently 4-parametric set of self-adjoint extensions \cite{math_book_selfadjext,math_deltalbergestexactsolv}. These extensions correspond to the discontinuities in wave function and its derivative. One of these extension is the case of $\delta$-interaction and allows implicit description of the regularization via smooth finite potentials. For other extensions where the wave function and its derivative are discontinuous the physical interpretations are not so explicit. They are the subject of discussion in many works (see \cite{qm_deltaresonant_jphys2003,qm_deltadipole_jmathphys2006,qm_deltaresonant_jmathphys2007}) starting from the work of ${\rm \check{S}}$eba \cite{qm_pointlike_repmathphys1986} where it was proved that no nontrivial $\delta'$ interactions exist in a sense of distributions over the space of continuous functions. The main result stated that these extensions corresponds to either completely disjoint semi-axes or well-known case of  $\delta$-interaction.

The problem of rigorous description of the singular point interactions  was solved by P.~Kurasov in \cite{funcan_deltadistr_jmathan1996,funcan_deltafinitrank_procmath1998} (see also monograph \cite{funcan_alberverio_singperturb}) from the point of view of theory of distributions.  These works, in particular, provided the correspondence between the self-adjoint extensions (i.e. boundary conditions) of \eqref{freeham} and its singular perturbations of the form:
\begin{equation}\label{kurasov_d2}
-D_{x}^2
\left(\, 1 + X_4\,\delta_{0} \,\right) + i\,D_{x}
\left(\, 2\,X_3\,\delta_{0} - i\,X_4\,\delta^{(1)}_{0}\right) + X_1\,\delta_{0}+(X_2 - i\,X_3)\,\delta^{(1)}_{0}\,.
\end{equation}
Here symbol $D_x$ stands for the derivative in the sense of distributions on the space of functions discontinuous at the point of singularity \cite{funcan_deltadistr_jmathan1996,funcan_deltafinitrank_procmath1998}:
\begin{equation}\label{kurasov_delta}
  \delta_{0}[\varphi] = \f{\varphi(0+0)+\varphi(0-0)}{2}\,,\quad
   \delta_{0}^{(1)}[\varphi] = \f{\varphi'(0+0)+\varphi'(0-0)}{2}\,\,.
\end{equation}
The parameters $X_i$ determine the values of the discontinuities of the wave function and its first derivative.

The aim of this paper is to give consistent physical interpretations to self-adjoint extensions of the free hamiltonian \eqref{freeham} which mimic point-like defects of different natures. We show that they fall into two classes: one of which is for pure potential perturbations and the other is for those which contain localized magnetic flux. From physical point of view the description of the localized defect in atomic chain by singular point-like interaction of free hamiltonian is based on junction of two boundary problems for corresponding semi-axes $\mathbb{R}_{-} = \{x<0\}$ and $\mathbb{R}_{+} = \{x>0\}$. One may expect that two classes of point-like defects are possible from the physical point of view. The first one is the class of the potential interactions which model the interaction of electrostatic nature. The magnetic interactions fall in the second class. The representation of the problem as the junction of two semi-axes allows to add symmetry consideration to the classification. The discontinuous action of the scaling symmetry can be interpreted in terms of the mass jump \cite{qm_deltamassjump_physlett2007,qm_deltamassjump_jphysmath2009}.
The electromagnetic gauge symmetry is related with the localized magnetic flux. Such classification in essential coincides with the results of symmetry analysis of work  \cite{funcan_deltasymm_lettmathphys1998}. We propose to identify different self-adjoint singular extensions of \eqref{freeham} with the continuous one-parameter subgroups of the gauge symmetry group. We give the corresponding relations of physical parameters with the mathematical parametrization used in \cite{funcan_deltasymm_lettmathphys1998}.

The structure of the paper is as follows. In Section~\ref{sec_gensymm} we give the basic symmetry consideration of the gauge symmetry for the free Hamiltonian on the real axis. In Section~\ref{sec_bcmagnetic} consider the regularization for the self-adjoint extension of localized magnetic flux. Section~\ref{sec_bcsimplectic} we apply symmetry consideration of Section~\ref{sec_gensymm} to the study of 4-parametric set of self-adjoint extensions and derive the relation between the mass-jump parameter with the parameter of $\delta^{(1)}_{0}$-interaction using the results of \cite{funcan_deltadistr_jmathan1996,qm_deltamassjump_jphysmath2009}.
Final remarks on the possibility of physical realization are in concluding section.

\section{Symmetry analysis of self-adjoint extensions of one-dimensional free Hamiltonian}\label{sec_gensymm}

The point-like interaction in 1D case means the joining the dynamics determined by the free particle Hamiltonian \eqref{freeham} on two semi-axes $\mathbb{R}_{-}$ and $\mathbb{R}_{+}$. Self-adjoint boundary conditions determine the unitary evolution operator on the whole real axis $\mathbb{R}$.

The Hamiltonian \eqref{freeham} allows two gauge transformations: the scaling $x \to \lambda\, x$ and gauge symmetry group $\mathcal{A}$ of the electromagnetic potentials (for simplicity we take $\mathcal{A}$ as constant). The group $\mathcal{G} = D \times \mathcal{A}$, where $D$ is the multiplicative group of positive real numbers and $\mathcal{A}$ is the additive group of gauge transformations. The unit element of $\mathcal{G}$ is $(1,0)$. This group may act either globally on the real axis $\mathbb{R}$ or disjointly on each of the semi-axes $\mathbb{R}_{-}\cup\mathbb{R}_{+}$. Indeed, one is free to choose both the scales for the coordinate $x$ and vector potential $A$ in corresponding semi-axes. All formal hamiltonians:
\begin{equation}\label{ham_gauge}
\hat{H} =  - \f{1}{2\,m}\left(\, -i\,\frac{d}{d\,x} - A \,\right)^2
\end{equation}
are gauge equivalent to the free particle Hamiltonian \eqref{freeham} provided that the region is connected. Joining the disconnected regions with different gauges results in appearance the gauge potential jumps.
In this sense the Hamitonian \eqref{freeham} is gauge equivalent to the Hamiltonian $\hat{H}_{\lambda} = \lambda^{-2}\,\hat{H}_{0}$. From this point of view the point-like singular self-adjoint extensions of the Hamiltonian \eqref{freeham} are the \textit{factor-Hamiltonians} representing the classes of the hamiltonians with the mass jump and the magnetic flux localized at the origin \cite{qm_deltamassjump_physlett2007}:
\begin{equation}\label{hamfreemassjump}
  \hat{H} =
  \begin{cases}
    -\f{1}{2\,m_{-}}\,
\left(\, -i\,\f{d}{d\,x}-A_{-} \,\right)
    \,, & \text{if}\,\,x<0 \\
    -\f{1}{2\,m_{+}}\,\left(\, -i\,\f{d}{d\,x}-A_{+} \,\right)\,, & \text{if}\,\,x>0\,.
  \end{cases}
\end{equation}
In other words the gauge symmetry can be broken at the junction. Thus the localized point-like defect of corresponding field appears.

The speculations given above demonstrate that the point-like
defects can be classified with respect to the action of these two gauge transformations (scaling and electromagnetic gauge potential $A$) acting separately on disjoint semi-axes. This action can be either continuous (global action on $\mathbb{R}_{-}\cup\mathbb{R}_{+}$), i.e. without mass and magnetic field jump or discontinuous with respect to these gauges. The continuous action means that $T^{-1}_{-}\,T_{+} = (1,0), T_{\pm}\in D \times \mathcal{A}$ and corresponds to the $\delta$-singularity. Other three cases are classified by the jump element $T^{-1}_{-}\,T_{+} = (\lambda_{+}/\lambda_{-}, A_{+} - A_{-})$ and the specific symmetry which is broken. The case $T^{-1}_{-}\,T_{+} =(1,A_{+} - A_{-})$ is naturally identified with the magnetic point-like flux which we consider via explicit regularization in the following Section.

\section{The boundary condition for localized magnetic flux}
\label{sec_bcmagnetic}

The junction of two semi-axes with discontinuity in value of the gauge potential $A$ at $x=0$ ($A(0+0)\ne A(0-0)$) means that at $x=0$ the localized flux exists. Such magnetic zero-range potential along with the corresponding boundary conditions can be easily represented as the limiting case of the proper regularization.

We consider the steady state of a particle in a magnetic field in the form
\begin{equation}
    \mathcal{H}(x)=
    \begin{cases}
        \mathcal{H}_{0}, & x\in[0,a]\\
        0, & x\not\in[0,a]
    \end{cases}
\end{equation}
where $a$ is of order of cyclotron radius. Hamiltonian in a magnetic field is as follows ($\hbar=1,e=1, c=1$):
\begin{equation}\label{ham}
    \hat{H}=\frac{1}{2m}\left(\,\hat{\mathbf{p}}-\mathbf{A}\,\right)^{2}\,.
\end{equation}
It is convenient to choose vector potential $\mathbf{A}$  in Landau gauge:
\[\mathbf{A}(x)=\mathcal{H}_{0}\,x\,\mathbf{e}_{y}\,.\]
In fact we consider 2D system with translational symmetry along $y$-axis. Since the Hamiltonian does not contain in an explicit form the coordinates $y$ then it commutes with the operator $\hat{p}_{y}$. From this it follows $y$ component of the momentum is conserved and accordingly choose $\psi$ in the form:
\begin{equation}\label{psi}
    \psi=e^{i\varphi(y)}\chi(x)\,.
\end{equation}
Substituting \eqref{psi} into \eqref{ham}, we obtain the following equation
\begin{equation}
       \tilde{ \chi}''(\tilde{x})+\left(\epsilon -(2\,\alpha\,\tilde{x}-\tilde{x}_{0})^2\right)\tilde{\chi}(\tilde{x})=0
\end{equation}
where $\tilde{x}=x\,k_x$, $\tilde{y}=y\,k_y$, $\epsilon=2\,m\,E\,k_x^2$, $\alpha=\Phi/\Phi_{0}$, $\Phi_{0}=2\,\pi\,\f{\hbar\,c}{q}$, $\tilde{x}_0=\varphi'(\tilde{y})\,k_y/k_x$.\\
Let us consider how the effect of a magnetic field $\mathcal{H}_{0}$ on the phase of the wave function. For this purpose we can calculate the gauge invariant phase difference between two points (with coordinates $\tilde{y}$ and $\tilde{y }+ d\tilde{y}$) by analogy with the way it is done for the Josephson junctions\cite{}:
\begin{equation}
    \varphi(\tilde{y }+ d\tilde{y})-\varphi(\tilde{y })=\oint\mathbf{A}\,\mathbf{d\,l}\,.
\end{equation}
The contour integral transforms into the flux of the magnetic field:
\begin{equation}
    \varphi-\varphi_{0}= \Phi \quad \text{or} \quad  \varphi=\varphi_{0}+2\pi\alpha\,,
\end{equation}
%
Thus
\begin{equation}\label{eq_eigen}
    \begin{cases}
       \tilde{ \chi}''(\tilde{x})+\left(\epsilon -4\,(\alpha\,\tilde{x})^2\right)\tilde{\chi}(\tilde{x})=0, & \tilde{x}\in[0,k_x\,a]\\
        \chi''(\tilde{x})+\epsilon \chi(\tilde{x})=0, & \tilde{x}\not\in[0,k_x\,a]
    \end{cases}
\end{equation}
Solution of the scattering problem \eqref{eq_eigen} is:
\begin{equation}
    \psi(\tilde{x})=
    \begin{cases}
        \exp(\imath\,\varkappa\,\tilde{x})+A\exp(-\imath\,\varkappa\,\tilde{x}), &\tilde{x}\leq0\\
        \exp(\imath\,\varphi)\left(B\,D_{-\frac{2\alpha-\epsilon}{4\,\alpha}}\left(2\,\tilde{x}\sqrt{\alpha}\right)+
        \tilde{B}\,D_{-\frac{2\alpha+\epsilon}{4\alpha}}\left(2\,i\,\tilde{x}\sqrt{\alpha}\right)\right), & \tilde{x}\in[0,a]\\
        C\,\exp(\imath\,\varkappa\,\tilde{x}), & \tilde{x}\geq\,a
    \end{cases}
\end{equation}
where $\varkappa=\sqrt{\epsilon}$.

The constraints of continuity of the function and its derivative on the boundary regions lead to a system:
\begin{equation}\label{bound_cond_ato0}
    \begin{cases}
        A+1=e^{i\phi_0}\left(\frac{B\left(\sqrt{\pi}\,
        2^{-\frac{2\alpha-\epsilon}{8\alpha}}\right)}{\Gamma\left(\frac{3}{4}-\frac{\epsilon}{8\alpha}\right)}+\frac{\tilde{B}\left(\sqrt{\pi}\,
        2^{-\frac{2\alpha+\epsilon}{8\alpha}}\right)}{\Gamma\left(\frac{3}{4}+\frac{\epsilon}{8\alpha}\right)}\right);\\
        i\,\varkappa(1-A)=-e^{i\phi_0}\left(\frac{B\left(\sqrt{\pi\alpha}\,2^{\frac{5}{4}+\frac{\epsilon}{8\alpha}}\right)}{\Gamma\left(\frac{1}{4}-\frac{\epsilon }{8\alpha}\right)}+\frac{\tilde{B}\left(i\sqrt{\pi\alpha}\, 2^{\frac{5}{4}-\frac{\epsilon}{8\alpha}}\right)}{\Gamma\left(\frac{1}{4}+\frac{\epsilon }{8\alpha}\right)}\right);\\
        e^{i(2\pi\alpha+\phi_0)}\left(B\,D_{-\frac{2\alpha-\epsilon}{4\alpha}}\left(2\,a\,\sqrt{\alpha}\right)+\tilde{B}\,D_{-\frac{2\alpha+\epsilon}{4\alpha}}\left(2\,i\,a\,\sqrt{\alpha}\right)\right)=
        C\,e^{i\,\varkappa\,a};\\
        e^{i(2\pi\alpha+\phi_0)}2\,\sqrt{\alpha}\,\left[B\,\left(a\,\sqrt{\alpha}\,D_{-\frac{2\alpha-\epsilon}{4\alpha}}\left(2\,a\,\sqrt{\alpha}\right)-D_{\frac{2\alpha+\epsilon}{4\alpha}}
        \left(2\,a\,\sqrt{\alpha}\right)\right)-\right.\\
        \left.-\tilde{B}\left(a\,\sqrt{\alpha}\,D_{-\frac{2\alpha+\epsilon}{4\alpha}}\left(2\,i\,a\,\sqrt{\alpha}\right)+i\,D_{\frac{2\alpha-\epsilon}{4\alpha}}
        \left(2\,i\,a\,\sqrt{\alpha}\right)\right)\right]=i\,\varkappa\,C\,e^{i\,\varkappa\,a}
    \end{cases}
\end{equation}
Solution of the system equation \eqref{bound_cond_ato0} in limit $a\to 0$ leads to the fact that wave function and its derivative are discontinuous
\begin{equation}\label{bound_cond}
    \begin{cases}
        \frac{\psi(0+0)}{\psi(0-0)}=e^{2\,i\,\pi\alpha}\\
        \frac{ \psi'(0+0)}{\psi'(0-0)}=e^{2\,i\,\pi\alpha}
    \end{cases}
\end{equation}
This is exactly one of the boundary conditions which makes the free hamiltonian \eqref{freeham} self-adjoint \cite{funcan_deltadistr_jmathan1996}. This case in Kurasov's formula \eqref{kurasov_d2} corresponds to the case $X_{1} = X_{2} = X_{4}=0\,,X_{3}\ne 0$ with:
\begin{equation}\label{magnet_x3alpha}
 e^{2\pi\,i\,\alpha} = \frac{2+i\,X_3}{2-i\,X_3}
\end{equation}
so the physical parameter $\alpha$, i.e. the magnetic flux is related with the mathematical parameter of the gauge field $X_3$ of \cite{funcan_deltadistr_jmathan1996}. Note that the magnetic (pseudovector) character of this extension also follows from the time reversal symmetry: $\psi\to \psi^*, X_3\to -X_3$. So from the physical pint of view it is quite clear that the regularized Schr\"{o}dinger operator for this case is \cite{funcan_deltadistr_jmathan1996}:
\begin{equation}\label{magnetic_hamkurasov}
  \hat{H} = (-i\,D_{x}- X_{3}\,\delta)^2 - (X_{3}\,\delta)^2\,.
\end{equation}
Note that $X_3$ does not necessary mean that $\alpha = 0$, rather the flux is quantized $\Phi/\Phi_0 = n$ in such a case.
\section{General form of boundary conditions}\label{sec_bcsimplectic}

From the standpoint of the theory of self-adjoint extensions Schr\"{o}dinger operators the boundary conditions for the singular one-point perturbation of the Hamiltonian \eqref{freeham} can be written in a form:
\begin{equation}\label{bound_cond2}
    \Gamma(0+0)=M\Gamma(0-0)\,,
\end{equation}
where
\begin{equation}
    \Gamma(0\pm0)=
    \left(
        \begin{array}{c}
            \psi(0\pm0) \\
            \psi'(0\pm0)
        \end{array}
    \right)\,,
\end{equation}
is the vector of the boundary values of a wave function and its derivative. We use the fact that the current density has the representation in terms of simplectic structure:
\begin{equation}
    j=\frac{1}{2\,i}\Gamma^{\dag}Sp_2\Gamma
\end{equation}
defined by the standard skew-symmetric matrix
\[Sp_2=
    \left(
        \begin{array}{cc}
            0 & 1 \\
            -1 & 0
        \end{array}
    \right)\,.
\]
From this it is easy to derive that the matrix $M$ satisfies
\begin{equation}\label{Sp_2}
    Sp_2=M^{\dag}Sp_2\,M\,,
\end{equation}
which means that it is the element of simplectic unitary group. These matrices can be parametrized in the following way:
\begin{equation}\label{m_param}
  M = z\,\left(
        \begin{array}{cc}
          1 & \frac{1}{(y-x)\,|z|^2} \\
          x & \frac{y}{(y-x)\,|z|^2} \\
        \end{array}
      \right)\,,
\end{equation}
where $x,y \in \mathbb{R}\,,\, x\ne y$ and $z \in \mathbb{C}$. In such parametrization the case of $\delta$-interaction is given by:
\begin{equation}\label{bc_delta}
  M_{\delta} = \left(
        \begin{array}{cc}
          1 & 0 \\
          X_1 & 1 \\
        \end{array}
      \right)
\end{equation}
with $y=\infty,\,x=X_1\,,\,z=1$. The matrix for the localized magnetic flux considered in Section~\ref{sec_bcmagnetic} is as following:
\begin{equation}\label{bc_magnetic}
    M_{\alpha}=e^{2\,i\,\pi\alpha}\,\left(
                             \begin{array}{cc}
                               1 & 0 \\
                               0 & 1 \\
                             \end{array}
                           \right)
\end{equation}
and corresponds to $x=0,\,y=\infty\,,\, z = e^{2\,i\,\pi\,\alpha}$. Within the symmetry classification noted above this is the case with the element $T^{-1}_{-}\,T_{+} = (1, \alpha)$.

Other two self-adjoint extensions correspond to the scaling symmetry and from physical point of view can be interpreted as the presence of mass jump~\footnote{authors indebted to Prof. V.~M.~Adamyan for this comment}.
One of these extensions is of potential nature and corresponds to the gauge subgroup $T^{-1}_{-}\,T_{+} = (q, 0)$. This extension is given by:
\begin{equation}\label{bc_delta'}
  M_{\delta'} = \left(
            \begin{array}{cc}
                1 &  - X_{4}\\
                0 & 1
            \end{array}
        \right)\,\,,
\end{equation}
\[ y=-1/X_4,\,\,x=0\,,\,\,z=1\,.\]
Due to its additive structure $M_{\delta'}(X_4)\, M_{\delta'}(X'_4) = M_{\delta'}(X_4+X'_4)$ it is
analogous to $\delta$-potential in this respect. This case is traditionally called $\delta'$-interaction (though it does not mean that $\delta'$ is added in any way to the hamiltonian according to \cite{qm_pointlike_repmathphys1986}).
According to \cite{funcan_deltadistr_jmathan1996} the Hamiltonian for \eqref{sdx4} is:
\begin{equation}\label{kurasov_x4}
\hat{H}_{\delta'} = -D^2_{x} \left(\, 1 + X_4\,\delta_{0} \,\right) + X_4\,D_{x}\,\delta^{(1)}_{0}\,.
\end{equation}
We state that it can be related with the Hamiltonian with the mass jump via regularization \cite{funcan_deltadistr_jmathan1996}. The result of \cite{qm_massjumphetero_prb1995} states the equivalence of the Hamiltonian with the spatially dependent mass $m(x)$ to the Hamiltonian with the constant mass and additional effective potential. This allows us to conclude about potential nature of this case.

Let us consider the case \eqref{bc_delta1}. In terms of \cite{funcan_deltadistr_jmathan1996} the boundary conditions for the case of so called $\delta^{(1)}$-potential is determined by the matrix:
\begin{equation}\label{bc_delta1}
  M_{\delta^{(1)}} = \left(\begin{array}{cc}
                \f{2+X_2}{2-X_2} & 0\\
                0 & \f{2-X_2}{2+X_2}
            \end{array}
        \right)\,\,,
\end{equation}
with $x=0,\,y=\infty$ and $z = \f{2+X_2}{2-X_2}$.
This extension was considered in \cite{qm_deltamassjump_jphysmath2009} in a situation where the mass jump $\mu = m_{+}/m_{-} \ne 1$ is present. The authors has shown that the following hamiltonian (we put $m_{-}=1$ by appropriately choosing the mass unit):
\begin{equation}\label{ham_gadella_mj}
  \hat{H} =
  \begin{cases}
    -\,
\f{d^2}{d\,x^2}
    \,, & \text{if}\,\,x<0 \\
    -\f{1}{\mu}\,\f{d^2}{d\,x^2}\,, & \text{if}\,\,x>0\,.
  \end{cases}
\end{equation}
has the self-adjoint extension with the boundary condition:
\begin{equation}\label{bc_gadella_mj}
     \left(
        \begin{array}{c}
            \psi(0+0) \\
            \psi'(0+0)
        \end{array}
    \right)
   = \left(
       \begin{array}{cc}
         \f{1+b}{1-\mu\,b} & 0 \\
         0 & \f{1-\,b}{1+\mu\,b} \\
       \end{array}
     \right)
    \left(
        \begin{array}{c}
            \psi(0-0) \\
            \psi'(0-0)
        \end{array}
    \right)\,,
\end{equation}
where the parameter $b$ is as follows \cite{qm_deltamassjump_jphysmath2009}:
\begin{equation}\label{b_gadella_mj}
  b = \f{1}{\sqrt{1+\mu+\mu^2}}\,,\quad \mu = \f{m_{+}}{m_{-}}\ne 1\,.
\end{equation}

As has been noted in \cite{qm_deltamassjump_jphysmath2009} if the mass jump is absent i.e. $\mu=1$, then \eqref{b_gadella_mj} is not valid and $b$ is free parameter of the extension ($b=X_2/2$). Thus the extensions with $\mu=1$ and $\mu\ne 1$ were treated as different ones. Here we show that there is the relation between the slef-adjoint extension \eqref{bc_delta1} of the free hamiltonian of constat mass and the self-adjoint extension of the hamiltonian \eqref{ham_gadella_mj} with mass jump.
Indeed, in accordance with the scaling property one can change the scale $x\to \lambda \,x$ for $x>0$ with $\lambda = 1/\sqrt{\mu}$ so that \eqref{ham_gadella_mj} transforms into \eqref{freeham}. Taking into account that $\psi_{0+0}\to \lambda^{-1/2}\,\psi_{0+0}\,,\,\,\psi'_{0+0}\to \lambda^{-3/2}\,\psi'_{0+0}$ we see that \eqref{bc_gadella_mj} goes to:
\begin{equation}\label{bc_gadella_scaled}
     \left(
        \begin{array}{c}
            \psi(0+0) \\
            \psi'(0+0)
        \end{array}
    \right)
   = \left(
       \begin{array}{cc}
         \lambda^{1/2}\,\f{1+b}{1-\mu\,b} & 0 \\
         0 &\lambda^{3/2}\, \f{1-\,b}{1+\mu\,b} \\
       \end{array}
     \right)
    \left(
        \begin{array}{c}
            \psi(0-0) \\
            \psi'(0-0)
        \end{array}
    \right)\,.
\end{equation}
Finally, comparing \eqref{bc_gadella_scaled} with \eqref{bc_delta1} we obtain the relation:
\begin{equation}\label{kurasov_mj_relation}
  X_2 = 2\,\frac{1+\mu^{5/4}- \mu^{1/4}\,\sqrt{\mu^2+\mu+1} +\sqrt{\mu^2+\mu +1}}{1-\mu ^{5/4}+\mu^{1/4}\,\sqrt{\mu ^2+\mu +1} +\sqrt{\mu ^2+\mu +1}}\,,\quad \mu \ne 1\,.
\end{equation}
Naturally, the limiting cases $\mu \to 0$ and $\mu \to \infty$ ($X_2 = 2$) represent two disjoint semi-axes.

This result shows that there is 1-1 correspondence between singular extension \eqref{bc_delta1} of the free Hamiltonian \eqref{freeham} and the extension given by \eqref{bc_gadella_mj} for the Hamiltonian with the mass jump \eqref{ham_gadella_mj} considered in \cite{qm_deltamassjump_jphysmath2009}. Note that the importance of scaling symmetry for \eqref{bc_delta1} was stressed in \cite{funcan_deltasymm_lettmathphys1998}. But according to our symmetry classification this extension corresponds to the generating element $T^{-1}_{-}\,T_{+} = (q, 1)$. This points to the quantized magnetic flux which is ``unobservable`` for spinless Hamiltonians. Additionally, we note that despite explicitly additive form of the Hamiltonian:
\begin{equation}\label{ham_delta1}
  H_{\delta^{(1)}} = -D^2_{x} - X_2\,\delta^{(1)}_{0}\,\,,
\end{equation}
the multiplication of M-matrices is not additive in parameter $X_2$:
\begin{equation}\label{x2_nonaddit}
M_{\delta^{(1)}}(X_2)\,M_{\delta^{(1)}}(X'_2)\ne M_{\delta^{(1)}}(X_2+X'_2)\,,
\end{equation}
in contrast to the cases \eqref{bc_delta} and \eqref{bc_delta'}.
Therefore in order to reveal the ``hidden`` magnetic flux in this case we need to include spin into consideration and one should use the Pauli Hamiltonian instead of \eqref{freeham}  \cite{book_ll3_en}.  The continuity of the related probability current:
\begin{equation}\label{j_spin}
{\rm \mathbf{J}}_{w} = {\f{{\hbar}} {{m}}}{\rm
Im}\left( {\Psi^{\dag} \nabla \Psi} \right) - {\f{q}{{mc}}}{\rm
\mathbf{A}}\Psi^{\dag} \Psi + {\f{{\hbar }}{{2m}}}{\rm rot}\left({\Psi^{\dag} \mathbf{\sigma} \,\,\Psi}  \right)\,,
\end{equation}
is the natural demand for self-adjointness.
Here:
\begin{equation}\label{psi_spin}
  \Psi = \left(
           \begin{array}{c}
             \psi_{\uparrow} \\
              \psi_{\downarrow} \\
           \end{array}
         \right)\,,
\end{equation}
and $\psi_{\uparrow}, \psi_{\downarrow}$ is the wave functions of corresponding spin states.

Let us check that the continuity of the probability current \eqref{j_spin} is fulfilled for the boundary conditions \eqref{bc_magnetic} and \eqref{bc_delta1} and the spin current term (the last term in \eqref{j_spin}) does not break the self-adjointness~\footnote{The second term with electromagnetic potential $A=const$ can be omitted due to gauge invariance}. Indeed, for $y$-component of the current \eqref{j_spin} we have:
\begin{equation}\label{jy_spin}
\left.\frac{\partial }{\partial x} \Psi^{\dag}\sigma_{z}\Psi \right|_{0+0} =\left.\frac{\partial }{\partial x}  (|\psi_{\uparrow}|^2 - |\psi_{\downarrow}|^2)\right|_{0+0} = \left.\frac{\partial }{\partial x} \Psi^{\dag}\sigma_{z}\Psi \right|_{0-0}\,\,,
\end{equation}
and it is easy to check that under both \eqref{bc_magnetic} and \eqref{bc_delta1} the equality:
\begin{equation}\label{spintermy_delta1}
\left.\frac{\partial }{\partial x} |\psi_{i}|^2 \right|_{0+0} =
\psi^*_{i}(0+0)\, \frac{\partial }{\partial x} \psi_{i}(0+0) = \psi^*_{i}(0-0)\, \frac{\partial }{\partial x} \psi_{i}(0-0)\,,\quad i=\uparrow, \downarrow
\end{equation}
is fulfilled. Analogously, considering the $z$-component of \eqref{j_spin}:
\begin{equation}\label{jz_spin}
\left.\frac{\partial }{\partial x} \Psi^{\dag}\sigma_{y}\Psi \right|_{0+0}=\left.\frac{\partial }{\partial x}  (\psi^{*}_{\uparrow}\,\psi_{\downarrow} - \psi^{*}_{\downarrow}\,\psi_{\uparrow})\right|_{0+0}\,\,,
\end{equation}
we see that \eqref{bc_magnetic} and \eqref{bc_delta1} conserve the current. In contrast to this the extensions \eqref{bc_delta} and \eqref{bc_delta'} do not allow point interaction with the magnetic field for spin particle. This proves that the point-like extensions \eqref{bc_delta} and \eqref{bc_delta'} are of pure potential nature and non magnetic, while \eqref{bc_magnetic} and \eqref{bc_delta1} are magnetic though the last case includes the localized magnetic flux as a discrete parameter of the extension. The results are summarized in the Table~\ref{tab_ext}.
\begin{table}[hbt!]
\center
\begin{tabular}{|c|c|c|c|}
  \hline
  BC & M-matrix & 1-parametric group & Parameters \\
  \hline
  I & $\left(
        \begin{array}{cc}
          1 & 0 \\
          X_1 & 1 \\
        \end{array}
      \right)$ & (1,0) & $\delta$-potential \\
  \hline
  II & $\left(
        \begin{array}{cc}
          1 & - X_4 \\
          0 & 1 \\
        \end{array}
      \right)$ & $\mathbb{R}_{+}$ & mass jump \\
  \hline
  III & $\left(
        \begin{array}{cc}
          e^{i\,2\,\pi\,\alpha} & 0 \\
          0 & e^{i\,2\,\pi\,\alpha} \\
        \end{array}
      \right)$ & $U(1)$ & magnetic \\
  \hline
  IV & $\left(\begin{array}{cc}
                \f{2+X_2}{2-X_2} & 0\\
                0 & \f{2-X_2}{2+X_2}
            \end{array}
        \right)$ & $\mathbb{R}_{+}\times \mathbb{Z}$ & magnetic \& mass jump \\
  \hline
\end{tabular}
\caption{Results}\label{tab_ext}
\end{table}

It is interesting to note that the generators of the extension matrices $M_i$:
\begin{equation}\label{m_gnrtrs}
  \mathfrak{g}_{i} = \left.\frac{d\,M_{i}}{d\,X_{i}}\right|_{X_{i}=0}
\end{equation}
are orthogonal in standard hermitian metrics
\begin{equation}\label{otho}
{\rm Tr} \left(\,   \mathfrak{g}_{i}\,\mathfrak{g}^{\dag}_{j} \,\right) \sim \delta_{ij}\,\,.
\end{equation}
\newpage
Additional support for the group classification of the self-adjoint extensions is the analysis of the S-matrix. By its symmetry properties $S$-matrix differentiates between non-magnetic hamiltonians and those with the magnetic field. We use the standard notations (see e.g. \cite{qm_faddeevyakub_eng}) for 1-dimensional scattering problem:
\begin{align}\label{psi_scat}
  \psi_{1} =&
  \begin{cases}
    e^{i\,k\,x}+A_{+}\,e^{-i\,k\,x}\,, & x\in \mathbb{R}_{-}, \\
    B_{+}\,e^{i\,k\,x} & x\in \mathbb{R}_{+}.
  \end{cases}
   \\
  \psi_{2} =&   \begin{cases}
    B_{-}\,e^{i\,k\,x}\,, & x\in \mathbb{R}_{-}, \\
    e^{-i\,k\,x}+A_{-}\,e^{i\,k\,x} & x\in \mathbb{R}_{+}.
  \end{cases}
\end{align}
\begin{equation}\label{smatrix_standard}
  \hat{S}  = \left(
               \begin{array}{cc}
                 A_{+} & B_{+} \\
                 B_{-} & A_{-} \\
               \end{array}
             \right)\,.
\end{equation}
Besides the general property of unitarity $\hat{S}^{\dag} = \hat{S}^{-1}$, time reversal symmetry demands the symmetricity of the S-matrix \cite{qm_faddeevyakub_eng}.

The corresponding $S$-matrices can be obtained easily. For point-like interaction it is given by:
\begin{equation}\label{sd}
    \hat{S}_{\delta} =\f{1}{2 k + i\,X_1}
    \left(
        \begin{array}{cc}
           -i\,X_1 & 2\,k\\
           2\,k & -i\,X_1
        \end{array}
    \right)
\end{equation}
The case of $\delta'$-interaction leads to the following $S$-matrix:
\begin{equation}\label{sdx4}
    \hat{S}_{\delta'} = \frac{1}{2+i\,k\,X_4}
    \left(
        \begin{array}{cc}
            i\,k\,X_4 & 2\\
            2 & i\,k\,X_4
        \end{array}
    \right)
\end{equation}
Both these S-matrices satisfy the general symmetry property $S^{*}(k) = \hat{S}(-k)$ and therefore correspond to nonmagnetic hamiltonians (potential interactions).
The localized magnetic flux characterized by:
\begin{equation}\label{s_magnetic}
    \hat{S}_{\alpha} =
    \left(
        \begin{array}{cc}
          0  & e^{2\,i\,\pi\,\alpha}\\
            e^{-2\,i\,\pi\,\alpha} & 0
        \end{array}
    \right)\,,
\end{equation}
i.e. the localized flux is reflectionless. Generally, $\hat{S}^{T}_{\alpha}\ne \hat{S}_{\alpha}$ due to the presence of the magnetic field. But it should be noted that quantized magnetic flux $\alpha \in \mathbb{Z}$ does not show itself in any way since the spin is switched off.

For the boundary conditions given by \eqref{bc_delta1} the $S$-matrix reads as:
\begin{equation}\label{s_delta1}
    \hat{S}_{\delta^{(1)}} =
    \left(
        \begin{array}{cc}
           \cos \theta  & \sin \theta\\
          \sin \theta   & -\cos \theta
        \end{array}
    \right)\,
\end{equation}
with
\[\tan \theta/2 = \f{2+X_2}{2-X_2}\,.\]
Both \eqref{s_magnetic} and \eqref{s_delta1} are hermitian, traceless and do not depend on the energy. Though the S-matrix \eqref{s_delta1} is symmetric and the boundary conditions as well as the formal Hamiltonian is real, nevertheless, the integer flux is present in this case as it follows from the group theoretical classification above and supported by the consideration of the Pauli Hamiltonian \eqref{j_spin}.

%
\section*{Conclusion}

The main result of the paper is the complete classification of the singular self-adjoint extensions of point-like nature for the free Hamiltonian \eqref{freeham}. These extensions can be classified with respect to the action of the gauge transformations corresponding to the choice of the mass scale and the phase of the wave function (electromagnetic gauge). As a bonus we state the exact relation \eqref{kurasov_mj_relation} between the singular $\delta^{(1)}$-potential interaction in a sense of \cite{funcan_deltadistr_jmathan1996} and the self-adjoint extension for the Hamiltonian with the step-wise mass jump considered in \cite{qm_deltamassjump_jphysmath2009}. In conclusion we point out that the result about the structure of singular $\delta^{(1)}$-interaction can be checked experimentally. The realization of boundary condition \eqref{bc_delta1} includes the mass jump in combination with the quantized flux at the junction. Naturally this can be realized in Josephson junctions and other one dimensional heterogeneous structures where the magnetic flux can be controlled in transition layer.

\subsection*{Acknowledgements}
The authors thank Prof. Vadim Adamyan for clarifying discussions. V.K. acknowledges Konstantin Yun for partial financial support of the research.

%

\end{document}